\documentclass[journal]{IEEEtran}

\IEEEoverridecommandlockouts   

\usepackage{mathptmx}
\usepackage[scaled=0.92]{helvet}
\usepackage{courier}
\usepackage{graphicx} 
\usepackage{amsmath}
\usepackage{bm}
\usepackage{amsfonts}
\usepackage{mathrsfs}
\usepackage{enumerate}
\usepackage{subfigure}
\usepackage{cite}

\DeclareMathOperator*{\argmin}{arg\,min}

\begin{document}
%
\title{Improved Initialization for Nonlinear State-Space Modeling}




\author{Anna~Marconato, 
        Jonas~Sj\"oberg, Johan~Suykens,
        and~Johan~Schoukens
\thanks{A. Marconato and J. Schoukens are with Dept. ELEC,  Vrije Universiteit Brussel, Pleinlaan 2, 1050 Brussels, Belgium. Email: anna.marconato@vub.ac.be. J. Sj\"oberg is with the Dept. of Signals and Systems, Chalmers University of Technology, Sweden. J. Suykens is with Dept. ESAT-SCD-SISTA, Katholieke Universiteit Leuven, Belgium.
This article has appeared as: A. Marconato, J. Sjöberg, J. A. K. Suykens and J. Schoukens, "Improved Initialization for Nonlinear State-Space Modeling," in IEEE Transactions on Instrumentation and Measurement, vol. 63, no. 4, pp. 972-980, April 2014.
doi: 10.1109/TIM.2013.2283553}
}
%


\maketitle

\begin{abstract}
This paper discusses a novel initialization algorithm for the estimation of nonlinear state-space models. Good initial values for the model parameters are obtained by identifying separately the linear dynamics and the nonlinear terms in the model. In particular, the nonlinear dynamic problem is transformed into an approximate static formulation, and simple regression methods are applied to obtain the solution in a fast and efficient way. The proposed method is validated by means of two measurement examples: the Wiener-Hammerstein benchmark problem, and the identification of a crystal detector.

\end{abstract}


%
\IEEEpeerreviewmaketitle

\section{Introduction}
\label{sec1}

In the instrumentation and measurement community there is an ever increasing demand for good models. Measuring and modeling are in fact closely linked together, since modeling techniques rely on raw data to build an analytical description of the system from which those data are generated. One can think of a very well-known example in electrical engineering, Ohm's law; this simple model explains the relation between two measured quantities, the voltage difference at two points of a conductor and the current flowing between the two points.

Having a look at more recent applications, accurate models are needed to tackle many problems in electrical engineering: to describe the behavior of sensors \cite{And13}, to characterize $\Delta\Sigma$ modulators and study their stability \cite{Lot13}, to better understand the functioning of wireless transmitters \cite{Rav12}, for fault diagnosis in analog circuits \cite{Den12}, just to mention a few examples.

Moreover, good models are very important in all branches of engineering, since they can be used to test a device in the design phase, to develop control units for industrial processes, and in general to replace expensive and time-consuming experiments in the analysis and simulation of a given system.

Most real-life systems (like the examples given above) are, to a certain extent, nonlinear. In particular, to get a thorough description of a given system, one needs to be able to model the dynamics as well as the (static) nonlinear behavior. 

To address the challenge of identifying nonlinear dynamic systems, one can consider nonlinear state-space (NLSS) models (here expressed in the discrete time domain):
\begin{align}
x(t+1)&=f(x(t),u(t)) \label{eq1a}\\
y(t)&=g(x(t),u(t)) \label{eq1b}
\end{align}
where $u(t)\in \mathbb{R}^{n_u}$ and $y(t)\in \mathbb{R}^{n_y}$ are the given input and output signal vectors at time instant $t$, $x(t)\in \mathbb{R}^{n_x}$ is the unknown state vector of the system, and $f(\cdot)$ and $g(\cdot)$ are the nonlinear functions to be estimated. 

Among the different model classes that can be used, state-space models show some advantages: they are general model structures that allow one to naturally describe system dynamics, and they are particularly suited for multiple input multiple output (MIMO) systems. 

Although state-space models have been extensively studied and employed in the context of linear system identification \cite{Lju99},\cite{PinSch12}, the identification of NLSS models is a far more complex task. 
In particular, a difficulty that is encountered when estimating NLSS models relates to the nature of the optimization problem. Nonlinear functions $f$ and $g$ are in general characterized by a number of parameters that need to be optimized, by minimizing a given criterion. If the problem is nonlinear in the parameters, an iterative search for the cost function minimum is performed, e.g. using a Levenberg-Marquardt technique \cite{Lju99},\cite{PinSch12}. 

Therefore, as for all iterative nonlinear optimization schemes, good initial values for the model parameters are needed, so that the time to convergence is reduced, and the problem of getting stuck in bad local minima is avoided. A possibility, often used in practice, is to initialize the nonlinear model with a simple linear description. However, a linear initialization does not always allow one to converge to a good local minimum, and therefore more advanced techniques are needed.

The goal of this paper is to propose a nonlinear initialization algorithm for the estimation of NLSS models. 

As will be explained in more details in the next sections, good initial values are obtained by transforming the nonlinear dynamic identification problem in Eqs.~(\ref{eq1a}-\ref{eq1b}) into a general nonlinear regression problem:
\begin{equation}
z(t)=h(\xi(t))
\label{nlregr}
\end{equation}
where $\xi(t)$ and $z(t)$ are the input and output signals respectively, and $h(\cdot)$ is the nonlinear function to be estimated, which is in general expressed as a basis function expansion \cite{sjo95}. This setting encompasses many different choices for the nonlinear basis functions: polynomials, splines, radial basis functions, sigmoids, and so on. 

Note that, since by applying regression methods the initialization procedure is speeded up considerably, different model structures for $f$ and $g$ can be tested more efficiently.

The main contributions of this paper are:
\begin{itemize}
\item obtaining good starting values for the optimization of NLSS models;
\item identifying separately the linear dynamics and the nonlinear terms in the model;
\item cutting the recursion in Eq.~(\ref{eq1a}) to obtain an approximate (static) version of the estimation problem, to which efficient regression methods can be applied.
\end{itemize}

This paper is organized as follows. The considered problem is presented in more details in Section~\ref{sec2}, together with an overview on the state of the art.
The different steps of the proposed identification algorithm are described in Section~\ref{sec3}.
The proposed method is validated by means of two challenging measurement problems: the Wiener-Hammerstein benchmark problem, and the identification of a crystal detector. The first example will be considered to illustrate the different aspects of the proposed approach in Section~\ref{sec7}. The second real data problem is discussed in Section~\ref{seccrystal}, together with a comparison of the obtained results with the ones given by other state-of-the-art methods. Concluding remarks are provided in Section~\ref{sec9}.

\section{Problem description}
\label{sec2}

The nonlinear dynamics in Eqs.~(\ref{eq1a}-\ref{eq1b}) are assumed to be modeled as:
\begin{eqnarray}
f(x(t),u(t))&=&Ax(t)+Bu(t)+f_{NL}(x(t),u(t)) \label{sstot1}\\
g(x(t),u(t))&=&Cx(t)+Du(t)+g_{NL}(x(t),u(t)) \label{sstot2}
\end{eqnarray}
where $A\in \mathbb{R}^{{n_x}\times{n_x}}$, $B\in \mathbb{R}^{{n_x}\times{n_u}}$, $C\in \mathbb{R}^{{n_y}\times{n_x}}$, $D\in \mathbb{R}^{{n_y}\times{n_u}}$ and $f_{NL}(\cdot)$ and $g_{NL}(\cdot)$ have $n_x$ and $n_y$ outputs respectively.

In this way, once the Best Linear Approximation (BLA) is estimated (that is, once an initial estimate for the matrices $A$, $B$, $C$ and $D$ is obtained, see Section~\ref{sec3} for more details), only the deviation from the linear model needs to be modeled, which has been proposed by several authors, see e.g. \cite{sjo97} and \cite{pad10}. Note that the proposed approach target systems for which the dynamics can be captured by the BLA and systems that are assumed to have only one equilibrium point.

The separation between the identification of the linear dynamics on one hand, and the estimation of the nonlinear terms on the other hand allows one to employ \textit{ad hoc} techniques specifically conceived to address the two different tasks, and to combine the advantages of both worlds. In particular, available linear system identification techniques (in the time or in the frequency domain) are used to determine the BLA, and nonlinear regression algorithms are employed to model the static nonlinearities  \cite{annai2mtc12}.

An estimate of the (unknown) nonlinear state is first computed, in order to cut the recursion loop in Eq.~(\ref{eq1a}). Thanks to this step, the NLSS estimation problem is transformed into a nonlinear regression problem of the form (\ref{nlregr}). 

Note that typically the nonlinear function $h(\cdot)$ is represented in terms of basis functions, for which one can choose among a large variety of possibilities: polynomials, splines, piecewise linear functions, radial basis functions, sigmoids, and in general any kind of static nonlinear function that one can think of. The formulation of the problem that is proposed in this paper to obtain an initial estimate of the nonlinear terms $f_{NL}$ and $g_{NL}$ in Eqs.~(\ref{sstot1}-\ref{sstot2}) is general enough to encompass many model structures from statistical learning. Support Vector Machines (SVMs), Least Squares SVMs, and Neural Network (NN) paradigms such as radial basis function (RBF) networks, multilayer perceptrons (MLPs), Extreme Learning Machines (ELMs), and so on, are all examples included in this general model class.

In the two measurement examples discussed in this paper, MLPs are considered to estimate the nonlinear terms in the state-space model in order to provide a simple illustration of the proposed algorithm, but the same approach could be followed employing a different nonlinear model structure.

In the classic (nonlinear) system identification framework, on the basis of a set of $N$ input/output measurement data $\left\{u(t),y(t)\right\}_{t=1}^{N}$, one can build a model characterized by a vector of parameters $\theta$ to describe the behavior of the underlying system. The obtained model can then be used to predict the output values $\hat{y}(t,\theta)$. 
For the algorithm presented here, beside the parameters characterizing the linear part of the model, the parameters appearing in the basis function expansion of the nonlinear terms mentioned above need to be estimated. Therefore, here $\theta$ contains both sets of parameters.

Following the Least Squares approach, optimal values of $\theta$ are found that minimize a least squares cost function $V$, typically the mean square error of the modeled outputs 
with respect to the true output values:
\begin{equation*}
\theta_{opt}=\arg \min_{\theta} V(\theta)
\end{equation*}
where
\begin{equation*}
V(\theta)=\frac{1}{N}\sum_{t=1}^N (y(t)-\hat{y}(t,\theta))^2
\end{equation*}

Since in most situations the resulting problem is nonlinear in the parameters, a numerical optimization is needed.

A typical difficulty that is encountered when minimizing the cost function $V(\theta)$ is the presence of a number of local minima in which the search algorithm may get trapped.
Therefore, choosing the starting values for parameters $\theta$ represents a crucial issue, since the initialization step has a big impact both on the quality of the final solution and on the time required for convergence. 

The goal of this work is to obtain good initial values of the parameter vector $\theta$, by combining system identification techniques to model the dynamics of the system and regression methods to estimate the nonlinearities. In this way, when fitting the parameters of model (\ref{eq1a}-\ref{eq1b}), one hopes to end up in a good (local) minimum to increase the quality of the final solution.  

\subsection{Related work}

An example of NLSS models that have been successfully applied to identify several real-life systems is given by polynomial NLSS (PNLSS) models \cite{pad10}. In that approach, the nonlinear model is initialized simply by a linear model (BLA).

A different approach is followed in \cite{cod94}, where a NN architecture characterized by a state-space structure is proposed and compared with the system input-output structure. Sines and cosines are used as activation functions in the network.

State-space models parametrized as feedforward NNs with one hidden layer and $\tanh(\cdot)$ as activation function are presented in \cite{suy95} and \cite{suy96}. For these neural state-space models, prediction error learning algorithms are discussed. In particular, as an alternative to the Extended Kalman Filter, the Kalman gain is directly parametrized in order to obtain simpler expressions. In this way, the gradient of the cost function can be computed in a straightforward manner by following Narendra's sensitivity model approach. To deal with local minima problems two different heuristics are proposed: the initialization of nonlinear neural state-space models as linear state-space models, and learning complex models starting from less complex ones, by increasing the number of hidden neurons in the model.

In order to make the recurrent nature of the problem vanish, in \cite{ver04} a method based on kernel canonical correlation analysis (KCCA) is employed to build the state sequence. The obtained state is then used together with the available input-output data to estimate the state-space model. Thanks to a Least Squares SVMs approach, a form of regularization is embedded within KCCA.

The use of KCCA is suggested also in \cite{win09} for the identification of nonlinear MIMO Hammerstein-Wiener systems. A subspace identification technique is used to estimate the nonlinearities separately from system dynamics.

Other approaches for the identification of NLSS models include the method proposed in \cite{tob10}, where knowledge on the state sequence is assumed to be available and is used to build the data records.

The Expectation-Maximization (EM) algorithm for maximum likelihood parameter estimation is instead used in \cite{gha99}, where Extended Kalman Smoothing is employed in the expectation step to get an estimate of the state. 

More recently, a combination of multilayer perceptron NNs, the EM algorithm and particle smoothing is suggested in \cite{gor08} for joint parameter and state estimation. 

Other examples of techniques that employ the EM algorithm in a maximum likelihood framework, and that use particle smoothing are presented in \cite{lin10} and \cite{schon11}. The same authors propose in \cite{nin10} a different approach that makes use of Fisher's identity and particle smoothing for gradient computation. 

Gaussian processes are used in \cite{tur10} for inference and learning, again using EM.

\subsection{Advantages and novelty of the proposed method}

The identification method presented in this paper for the estimation of NLSS models, and more specifically the proposed initialization algorithm, shows a number of differences and novel aspects compared with the other approaches mentioned above:
\begin{itemize}
\item the separation between the linear dynamics and the static nonlinearities in the formulation of the problem allows one to identify the different parts independently, in a more effective way;
\item the initialization scheme is based on a combination of ideas coming from different worlds - system identification and nonlinear regression/statistical learning - exploiting the advantages of both;
\item the BLA is used to incorporate dynamics in methods that are essentially designed to model static nonlinearities;
\item the proposed algorithm is a general scheme that can be used in combination with many different choices of nonlinear model structures.
\end{itemize}

\section{Proposed algorithm}
\label{sec3}

In this section, the proposed method for the estimation of NLSS models is presented. First of all, the initialization algorithm is explained, then the next steps of the method are discussed, followed by the description of the considered model structure. Finally, some comments on time saving issues are provided.

\subsection{Initialization scheme}

The proposed scheme for the initialization of NLSS models consists of three main steps:
\begin{enumerate}
\item obtain a linear model to capture the dynamics of the system;
\item estimate the nonlinear state;
\item model the nonlinearities.
\end{enumerate}

In this section all the different steps are described in details.

{\par\ \par}

\subsubsection{Obtain a linear model}

First of all, the nonlinear input-output behavior is approximated with a linear model, by estimating the BLA \cite{PinSch12}. Among the possible choices of linear models that one can use, the BLA is defined to be optimal in least square sense. More in details, in the set of linear models $\mathscr{G}$, the BLA is defined as the model $G$ such that:
$$G_{BLA}=\argmin_{G\in\mathscr{G}}{\mathscr{E}\{\left|y(t)-G(u(t))\right|^2\}}$$ where $u(t)$ and $y(t)$ are the input and output of the nonlinear system and $\mathscr{E}$ is the expected value with respect to the input \cite{PinSch12},\cite{Enq05}. In this way matrices $\hat{A}$, $\hat{B}$, $\hat{C}$ and $\hat{D}$ can be determined, obtaining the following linear model:

\begin{align}
x(t+1)&=\hat{A}x(t)+\hat{B}u(t)\\
y(t)&=\hat{C}x(t)+\hat{D}u(t)
\end{align}

The linear model can then be used to get an approximation of the nonlinear state, as discussed in the next paragraph.

{\par\ \par}

\subsubsection{Estimate $\hat{x}_{LS}$}

A crucial aspect in the proposed approach is based on the fact that if the state $x(t)$ would be exactly known, the problem of obtaining a nonlinear model could be solved much more easily by estimating $f$ and $g$ individually and as static mappings. Since the nonlinear state is in practice not available, one would like to obtain an approximation of $x(t)$, to be able to obtain initial estimates of $f$ and $g$.

In particular, using the available data $\left\{u(t),y(t)\right\}_{t=1}^{N}$ and the BLA estimates $\hat{A}$, $\hat{B}$, $\hat{C}$, $\hat{D}$ obtained in the previous step, the nonlinear state is approximated as a trade-off between the linear model and the data fit, by solving the following Least Squares (LS) problem:

\begin{align}
\hat{x}_{LS}(t)=& \arg\min_{\{x(t)\}} \sum_t{(y(t)-\hat{C}x(t)-\hat{D}u(t))^2} \nonumber \\ & + \lambda \sum_t{(x(t+1)-\hat{A}x(t)-\hat{B}u(t))^2} \nonumber \\ = & \arg\min_{\{x(t)\}} E_y + \lambda E_x
\label{eqLS}
\end{align}

The first term $E_y$ of the cost function represents the data fit, while the second term $E_x$ represents the linear model fit; $\lambda$ is the trade-off parameter that needs to be tuned to change the emphasis given on the two criteria. By tuning $\lambda$ a deviation from the linear state (resulting from the BLA estimates $\hat{A}$, $\hat{B}$, $\hat{C}$, $\hat{D}$) is allowed, to take into account the nonlinear terms in Eqs.~(\ref{sstot1}-\ref{sstot2}). In practice, the optimal value for the trade-off parameter $\lambda$ is chosen from a grid of values as the one minimizing the root mean square error, RMSE, (on the validation data) of the obtained initialized nonlinear models. The tuning of the trade-off parameter $\lambda$ will be illustrated in more details on the Wiener-Hammerstein example in Section~\ref{sec7}.

Problem (\ref{eqLS}) could be replaced by a Kalman filter which would change the approximation slightly \cite{kalman}.

{\par\ \par}

\subsubsection{Estimate nonlinear functions $f$ and $g$}
\label{sec33}

Once the estimate $\{\hat{x}_{LS}(t)\}_{t=1}^{N}$ of the nonlinear state is available, one obtains the following approximate static problem:
\begin{eqnarray}
\hat{x}_{LS}(t+1)&=&f(\hat{x}_{LS}(t),u(t))+r_{LS}(t)= \hat{A}\hat{x}_{LS}(t)+\hat{B}u(t)+\nonumber \\ & &+f_{NL}(\hat{x}_{LS}(t),u(t))+r_{LS}(t) \label{sstot1appr}\\
y(t)&=&g(\hat{x}_{LS}(t),u(t))+e_{LS}(t)= \hat{C}\hat{x}_{LS}(t)+\hat{D}u(t)+\nonumber \\ & &+g_{NL}(\hat{x}_{LS}(t),u(t))+e_{LS}(t) \label{sstot2appr}
\end{eqnarray}
where $r_{LS}(t)$ and $e_{LS}(t)$ are error terms resulting from the fact that here the approximated nonlinear state is introduced in the problem.
Eqs.~(\ref{sstot1appr}-\ref{sstot2appr}) represent two static regression problems that can be solved independently employing simple regression methods. Note that at this stage the recursion in the state equation is not present anymore, since the state sequence is now assumed to be `known'.
Therefore, both functions $f_{NL}(\hat{x}_{LS}(t),u(t))$ and $g_{NL}(\hat{x}_{LS}(t),u(t))$ can be estimated as basis function expansions.

\subsection{Simulation of the initialized model and optimization}

The two estimated nonlinearities $\hat{f}_{NL}$ and $\hat{g}_{NL}$ can then be included in a general NLSS structure; at this point the dynamics are again taken into account, and one can simulate the obtained initialized nonlinear model to assess its performance. In other words, the recursion in the state equation is included again, switching back from the approximate initialization obtained using (\ref{sstot1appr}-\ref{sstot2appr}) to (\ref{sstot1}-\ref{sstot2}), which is the original problem one wants to solve.

Finally, the obtained initial estimate of the nonlinear model can be further fitted to data, using an iterative nonlinear optimization routine, e.g. the Levenberg-Marquardt algorithm.

\subsection{Model structure}

\begin{figure}[!bt]
\centering
\subfigure[]{
\includegraphics[width=6cm]{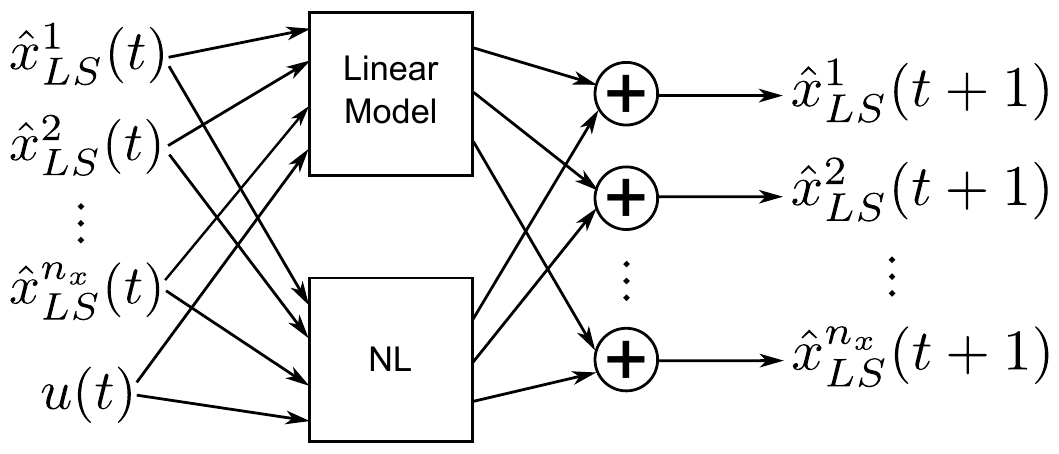}
\label{figNLf}
}
\subfigure[]{
\includegraphics[width=5cm]{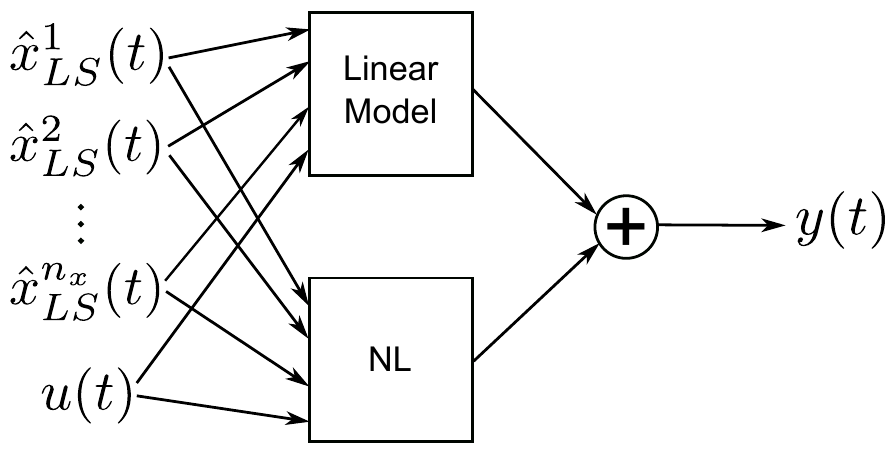}
\label{figNLg}
} 
\caption{Model structure used to describe (a) the state equation, (b) the output equation, in the case of one input, one output and $n_x$ states. The nonlinear block is added in parallel to the linear model.}
\label{figNL}
\end{figure}

To describe the NLSS model, the model structure represented in Fig.~\ref{figNL} is used. In Fig.~\ref{figNLf} the representation of state equation (\ref{eq1a}) is given in the case of having one input, one output and $n_x$ states. In the block scheme the inputs are the states $\hat{x}_{LS}^1(t), \hat{x}_{LS}^2(t), \ldots, \hat{x}_{LS}^{n_x}(t)$ and the input $u(t)$, while the outputs are the states $\hat{x}_{LS}^1(t+1), \hat{x}_{LS}^2(t+1), \ldots, \hat{x}_{LS}^{n_x}(t+1)$. The nonlinear block is added in parallel to the linear model, so that the update for the states consists of a linear plus a nonlinear part, see Eq.~(\ref{sstot1}). The same structure is used to describe the output equation (\ref{eq1b}) in Fig.~\ref{figNLg}. 

\subsection{Time saving}
\label{sec4}
This paper presents a nonlinear initialization algorithm for the estimation of NLSS models. Another possibility, which is often used in system identification, to initialize a nonlinear model before the optimization step is to use a simple linear initialization. Note that by `linear initialization' we mean that a nonlinear model is estimated by starting from a linear model in parallel with nonlinear terms that are initialized to zero; all parameters (of both the linear and the nonlinear part) are then optimized \cite{sjo97}.

As will be illustrated in the next section with the Wiener-Hammerstein example, a linear initialization does not always guarantee convergence to a good local minimum of the cost function. In most cases, when starting from a linear model, the optimization routine gets stuck in bad local minima. 

Moreover, by starting from better initial estimates, the time to convergence needed when fitting the model parameters will be reduced (or, dually, given the same amount of computational time the final fitted model is likely to be characterized by lower error values).

Beside this advantage, there are also other aspects of the initialization algorithm presented in this work that allow one to reduce the computational time.

Firstly, time can be saved since one can choose to perform model (structure) selection already after the initialization phase. This is obviously not the case if a linear initialization is applied, since all initialized linear models result in the same error value. This means that with the proposed scheme, if a number of the possible nonlinear models can be discarded after initialization, one would not need to optimize all initialized models to be able to determine their performance. This aspect will be discussed further in Section~\ref{sec7}.

Secondly, here the initial estimates of the nonlinear functions $f$ and $g$ are computed separately, i.e. if many nonlinear model structures need to be tried out this approach allows one to carry out all estimations in much less time. 

\section{Wiener-Hammerstein benchmark problem}
\label{sec7}

We illustrate the different aspects of the proposed initialization algorithm for NLSS models on the Wiener-Hammerstein benchmark example. One-hidden-layer MLPs with $\tanh(\cdot)$ as neuron activation function are chosen to model the nonlinear terms. The considered model structure is then the one depicted in Fig.~\ref{figNL}, where the two nonlinear blocks (NL) are two MLPs. 

During the training phase, the parameters of the MLP are typically first randomly initialized so that the active regions of the neurons span the input space, and then optimized, e.g. by means of the backpropagation algorithm \cite{has09}. Therefore, in the following many different initial MLP configurations are considered, to show also the effect of this variability source on the obtained model performance.

\subsection{Description of the data}

The data are generated from a nonlinear electronic system, characterized by a Wiener-Hammerstein structure.
The two linear blocks are a third order Chebyshev low-pass filter with 0.5 dB ripple and cut-off frequency at 4.4 kHz, and a third order inverse Chebyshev low-pass filter with a -40 dB stop band from 5 kHz, respectively. The static nonlinearity is built using two resistors and a diode.

The goal is to identify a model to describe this nonlinear dynamic system, based on a set of real input/output measurements. A difficulty that pops up in the identification of this system is given by the fact that it is not straightforward to model the nonlinear part of the system, since this is not directly accessible from either the input or the output.

The data were collected by exciting the system with a filtered Gaussian input signal (cut-off frequency at 10 kHz). The sampling frequency is equal to 51.2 kHz. A very large dataset is available\footnote{Download at http://tc.ifac-control.org/1/1/Data\%20Repository/sysid-2009-wiener-hammerstein-benchmark.}
 (188000 input/output samples) to estimate and validate the model, but in this work shorter records are considered to reduce the computational load. To illustrate the different aspects of the proposed algorithm, an estimation set of 2500 samples is considered to build the NLSS model, and a validation set of 10000 samples is used to evaluate the performance of the obtained model on fresh data. A very large validation set is chosen in order to eliminate the stochastic variations in this step, so that the model quality is clearly visible. In real life applications, one would instead choose to use a larger portion of the available data to estimate the model.

\subsection{Initialization}

\subsubsection{Estimate of the linear model}

The BLA is computed to obtain a first (linear) description of the system. A 6th order linear state-space model is obtained, characterized by a RMSE value of 60 mV on the validation set.

{\par\ \par}

\subsubsection{Estimate of the nonlinear state}

\begin{figure}[!t]
\centering
\includegraphics[width=8.0cm]{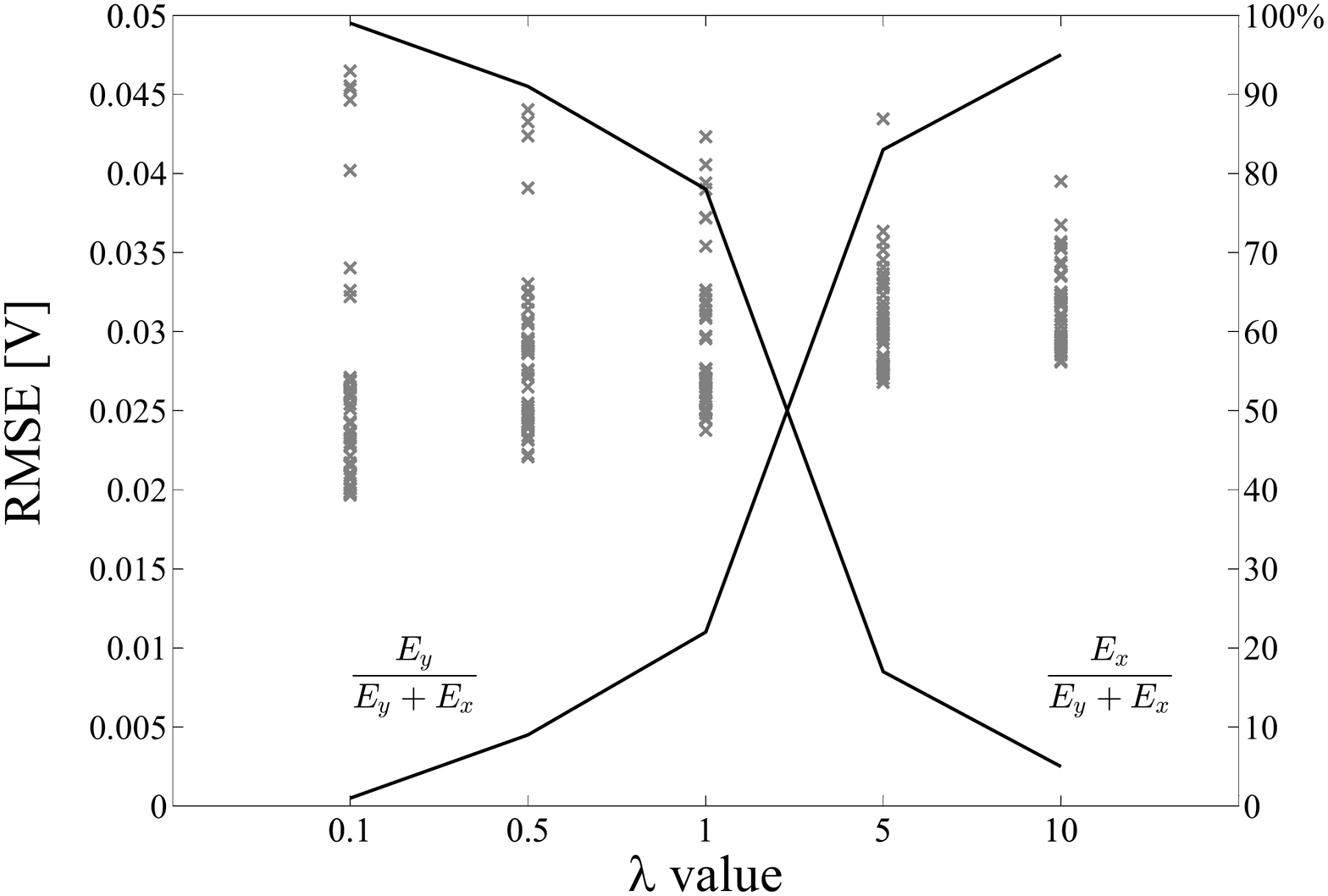}
\caption{Wiener-Hammerstein example. Initialized NLSS models (gray crosses): RMSE values on the validation set, for different values of $\lambda$ (50 initial MLP configurations, number of neurons for $f_{NL}$ and $g_{NL}$ equal to 3). The relative weight (in percentage) of the two error criteria $E_y$ and $E_x$ is also shown (right side axis) for the different $\lambda$ values.}
\label{figlambdainit}
\end{figure}

Eq.~(\ref{eqLS}) is used to estimate the nonlinear state sequence. The main issue here concerns the tuning of the trade-off parameter $\lambda$. First of all, one needs to determine a suitable range of values for $\lambda$. This is done by balancing the weight of the two criteria $E_y$ and $E_x$ in the trade-off, i.e. by choosing several values of $\lambda$ for which either $E_y$ or $E_x$ is dominant, or both are equally important. In this example, $\lambda$ is chosen among these values: $0.1, 0.5, 1, 5, 10$. The effect of changing $\lambda$ on the importance of the two criteria is shown in Fig.~\ref{figlambdainit}.
The choice is made for example by selecting the value for which the RMSE of the resulting initialized NLSS model on the validation set is minimized. 

The obtained results on the validation set are plotted in Fig.~\ref{figlambdainit}, for 50 different random initial configurations of the MLPs.

The lowest RMSE values are obtained for $\lambda=0.1$, so if one decides to tune $\lambda$ after the initialization phase, without optimizing all initialized models, this will be the chosen value. As will be shown later in the part dedicated to the optimization, the final performance of the optimized models does not seem to depend much on the specific value chosen for $\lambda$ (see Fig.~\ref{figlambdaopt}).

However, for each $\lambda$, the variability due to the different MLP configurations has a significant impact on the performance of the resulting NLSS models, so in general one might need to initialize a number of different models, and check their performance once all parameters have been optimized. Note that this issue would not be present if a different choice of the basis functions in the estimation of the nonlinear terms was considered.

{\par\ \par}

\subsubsection{Estimate of the nonlinear terms}

To complete the initialization step, the two nonlinear terms $f_{NL}$ and $g_{NL}$ needs to be estimated, by solving the static regression problems in Eqs.~(\ref{sstot1appr}-\ref{sstot2appr}). As already mentioned, MLPs are used here to model the two nonlinearities, so for both $f_{NL}$ and $g_{NL}$ the number of neurons in the hidden layer ($n_f$ and $n_g$ respectively) needs to be specified. Again, this is a model selection problem that can be solved at initialization. 

Table~\ref{neuronsinit} shows the RMSE (on the validation set) obtained by using different values for $n_f$ and $n_g$ (for the sake of simplicity, here $n_f=n_g$). For each number of neurons, 20 different MLP configurations are considered, and the lowest RMSE value is shown in the table, to give an indication of how low the error can be pushed with models of increasing complexity.

\begin{table}[!t]
\begin{center}
\begin{tabular}{| c | c | c | c | c |}
\hline
$n_f=n_g$ & 1 & 2 & 3 & 4 \\
\hline
RMSE [V] & 0.0410 & 0.0256 & 0.0197 & 0.0196 \\
\hline
\end{tabular}
\caption{\label{neuronsinit} {Wiener-Hammerstein example. Initialized NLSS models: lowest RMSE values obtained on the validation set for different values of $n_f$ and $n_g$ ($\lambda=0.1$).}}
\end{center}
\end{table}

At this stage, the model obtained with 3 neurons can be considered the best one, since the small gain in performance using 4 neurons does not justify the corresponding increase in the number of parameters. Table~\ref{neuronsopt} shows that the same reasoning holds also for the optimized models resulting from these initializations.

\subsection{Optimization}

\begin{figure}[!t]
\centering
\includegraphics[width=8.0cm]{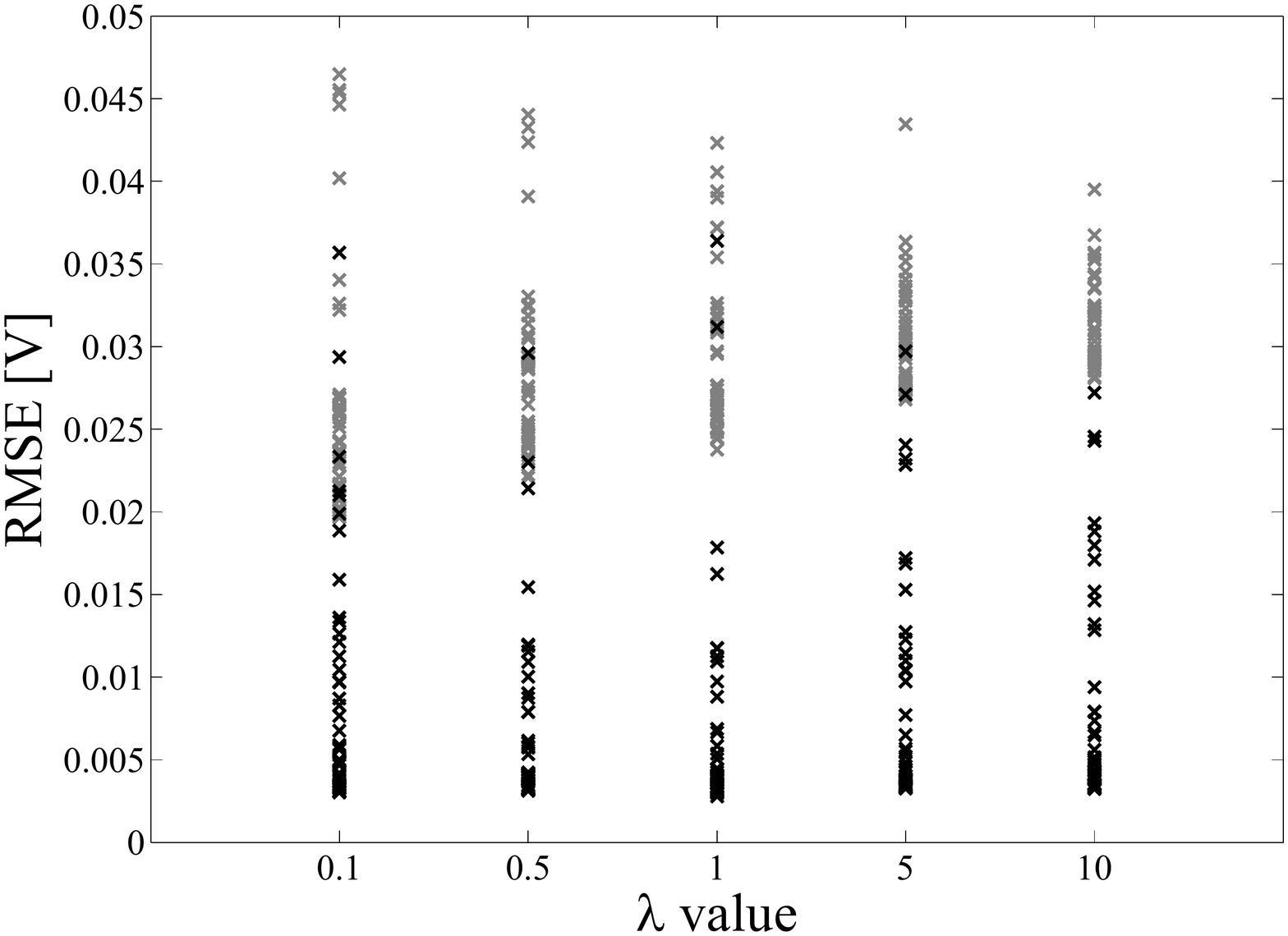}
\caption{Wiener-Hammerstein example. Optimized NLSS models (black crosses): RMSE values on the validation set, for different values of $\lambda$ (50 initial MLP configurations, number of neurons for $f_{NL}$ and $g_{NL}$ equal to 3). The results for the initialized models are repeated here for a comparison (gray crosses).}
\label{figlambdaopt}
\end{figure}

Although, as already pointed out, the choice of $\lambda$, $n_f$ and $n_g$ can be done after the initialization step, all the initialized models considered above have been optimized, to illustrate the main properties of the proposed algorithm. The obtained results are summarized in Fig.~\ref{figlambdaopt} and Table~\ref{neuronsopt}.

\begin{table}[!t]
\begin{center}
\begin{tabular}{| c | c | c | c | c |}
\hline
$n_f=n_g$ & 1 & 2 & 3 & 4 \\
\hline
RMSE [V] & 0.0129 & 0.0045 & 0.0034 & 0.0032 \\
\hline
\end{tabular}
\caption{\label{neuronsopt} {Wiener-Hammerstein example. Optimized NLSS models: RMSE values on the validation set after optimizing the 4 initialized models of Table~\ref{neuronsinit}, for different values of $n_f$ and $n_g$.}}
\end{center}
\end{table}

The main conclusion is that, for this example, the choice of $\lambda$ is not too critical, since it does not affect much the final performance of the optimized models. It is rather the variability due to the different MLP configurations that can yield models with a higher RMSE (Fig.~\ref{figlambdaopt}). 

However, all initialized models improve the result obtained by the BLA, and with the proposed initialization scheme it is possible to obtain (after optimization) NLSS models characterized by low RMSE and low complexity in terms of number of parameters. As for the computational aspects, in this example (run using MATLAB, on a PC with a i7-2600 3.40 GHz processor) the nonlinear initialization takes approximately 8 seconds, to be compared with a processing time of 40 minutes needed for the optimization routine. Therefore, the initialization step, even when repeated several times, can be considered inexpensive.

Note that similar conclusions can be drawn also when testing the proposed approach on examples where stronger nonlinearities (such as the absolute
value) are present in the system. In those cases, the only difficulty seems to be the fact that it is not always possible to perform model selection already at initialization, and one then needs to optimize several nonlinear models before selecting the optimal one.

\subsection{Comparison with linear initialization}

To demonstrate the advantage of using the proposed algorithm to initialize NLSS models, two situations are compared:
\begin{itemize}
\item Linear/Random initialization: the MLP parameters associated to the neuron amplitudes are set equal to 0, while all other MLP parameters (associated to the neuron positions) are randomly generated;
\item Linear/MLP initialization: the proposed initialization is performed (with $\lambda=0.1$) to determine the MLP parameters associated to the neuron positions, but the MLP parameters associated to the neuron amplitudes are set equal to 0.
\end{itemize}

For both initializations, 50 different MLP configurations are considered, with $n_f=n_g=3$. 

All models resulting from these two initializations are linear models (i.e. they have the same performance as the BLA), but in one case (Linear/MLP) one exploits the power of the proposed algorithm to fix the positions of the neurons. The mean value and the standard deviation (in V) of the RMSE results (on the validation set) obtained after optimizing all models are listed below:
\begin{itemize}
\item Linear/Random: mean $0.0443$, std $0.0008$;
\item Linear/MLP: mean $0.0068$, std $0.0054$.
\end{itemize}

This shows that if no information is given about the positions of the MLP neurons, the optimization routine gets always stuck in a bad local minimum. Only when the proposed algorithm is exploited to determine (at least) the neurons positions, it is possible to converge to lower RMSE values.

A final remark on the initialization: the algorithm discussed in this paper seems to be powerful (in comparison with a linear initialization) in those cases in which the basis functions describing the nonlinearities are characterized by parameters expressing the `position', i.e. like in the MLP example. For `position-free' models, e.g. polynomial models, a linear initialization might be effective enough for the optimization to converge to a good local minimum.

\section{Identification of a crystal detector}
\label{seccrystal}

In this section,
the task is to model the input/output relationship of a Agilent-HP420C crystal detector, 
a device that is often used in microwave applications to measure the envelope of a signal \cite{liesbeth07}.

\subsection{Description of the data}

The excitation signal consists of 5 measured periods of a Gaussian noise sequence (each period is made of 50000 samples, at a sampling frequency of 10 MHz) characterized by a slowly increasing amplitude. A first data set was generated by considering a bandwidth of 800 kHz and is used to estimate the model, while a second data set, used for validation purposes, was generated with a similar excitation signal with bandwidth equal to 400 kHz. The 5 different periods of the excitation signal are used in the first step of the initialization when estimating the BLA to reduce noise and transient effects, while later on, for the estimation of the nonlinear model, the averaged input signal (over 5 periods) is considered.

\begin{figure}[!t]
\centering
\includegraphics[width=6.0cm]{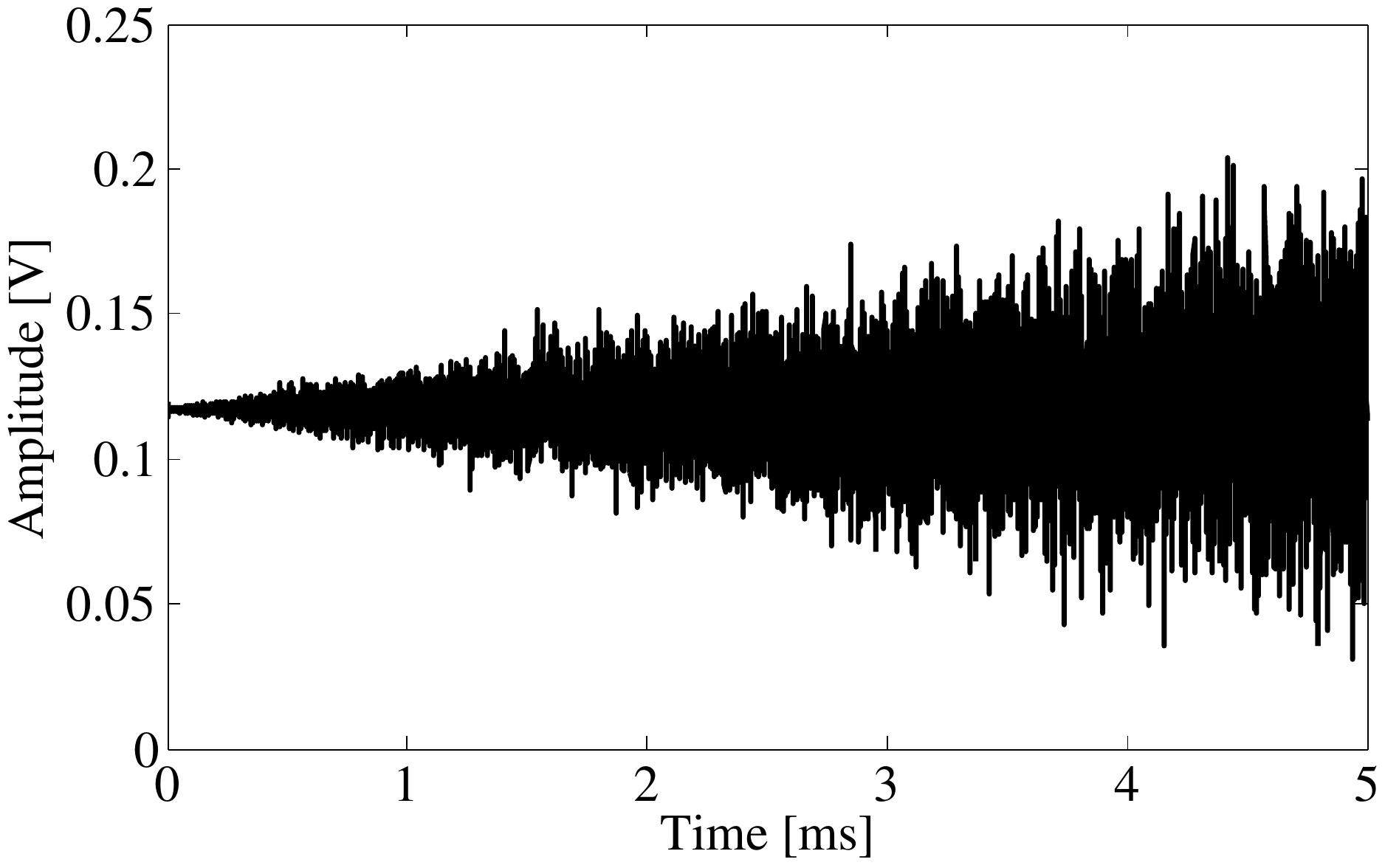}
\caption{Crystal detector example. Averaged input signal (for the estimation data set).}
\label{figinput}
\end{figure}

The input signal (averaged over 5 periods) used for estimation is shown in Fig.~\ref{figinput}.

The noise on the input and output measurements is estimated to have a standard deviation of 0.23 mV, to be compared to a RMS value of 15.2 mV for the input, and 13.6 mV for the output (after removing the DC offset).

\subsection{Obtained results}

For the initialization of the NLSS model, first of all the BLA is estimated nonparametrically to capture the dynamics of the system \cite{PinSch12}, then a first order parametric linear model is obtained and transformed into state-space form. The BLA results in a RMSE equal to 1.5 mV and 1.1 mV on the estimation and validation set, respectively. 

The nonlinear state $x_{LS}$ is then approximated solving the problem in Eq.~(\ref{eqLS}). Here the optimal value for the trade-off parameter $\lambda$ is chosen to be 0.01 (corresponding to $E_y/(E_y+E_x)$ less than $1\%$), although also in this example the choice of $\lambda$ does not seem to have a significant impact on the final result.

To estimate the nonlinear functions $f$ and $g$, the model structure in Fig.~\ref{figNL} is considered, with the difference that now only one state is present, since a first order model is sufficient to describe the linear dynamics. One-hidden-layer MLPs with $\tanh(\cdot)$ as activation function are used to estimate the nonlinear terms, and the choice of two neurons for $f_{NL}$ and one neuron for $g_{NL}$ results in an initialized nonlinear model that significantly improves the performance of the BLA (RMSE equal to 0.65 mV on the validation data). 

Starting from the initial estimates, all model parameters are then optimized using a Levenberg-Marquardt algorithm. This finally results in a fitted nonlinear model characterized by a RMSE equal to 0.27 mV on the validation set.

All obtained results in terms of RMSE on estimation and validation data are summarized in Table~\ref{tabcrystal}. 

\begin{table}[!t]
\begin{center}
\begin{tabular}{| c | c | c |}
\hline
\textbf{Model} & \textbf{Estimation RMSE} & \textbf{Validation RMSE}  \\
\hline
\hline
Best Linear Approximation & 1.50 & 1.10 \\
\hline
Initialized nonlinear model & 0.72 & 0.65 \\
\hline
Optimized nonlinear model & 0.26 & 0.27 \\
\hline
\end{tabular}
\caption{\label{tabcrystal} {Crystal detector example. RMSE values (in $mV$) obtained by the considered models.}}
\end{center}
\end{table}

\begin{figure}[!t]
\centering
\includegraphics[width=6.0cm]{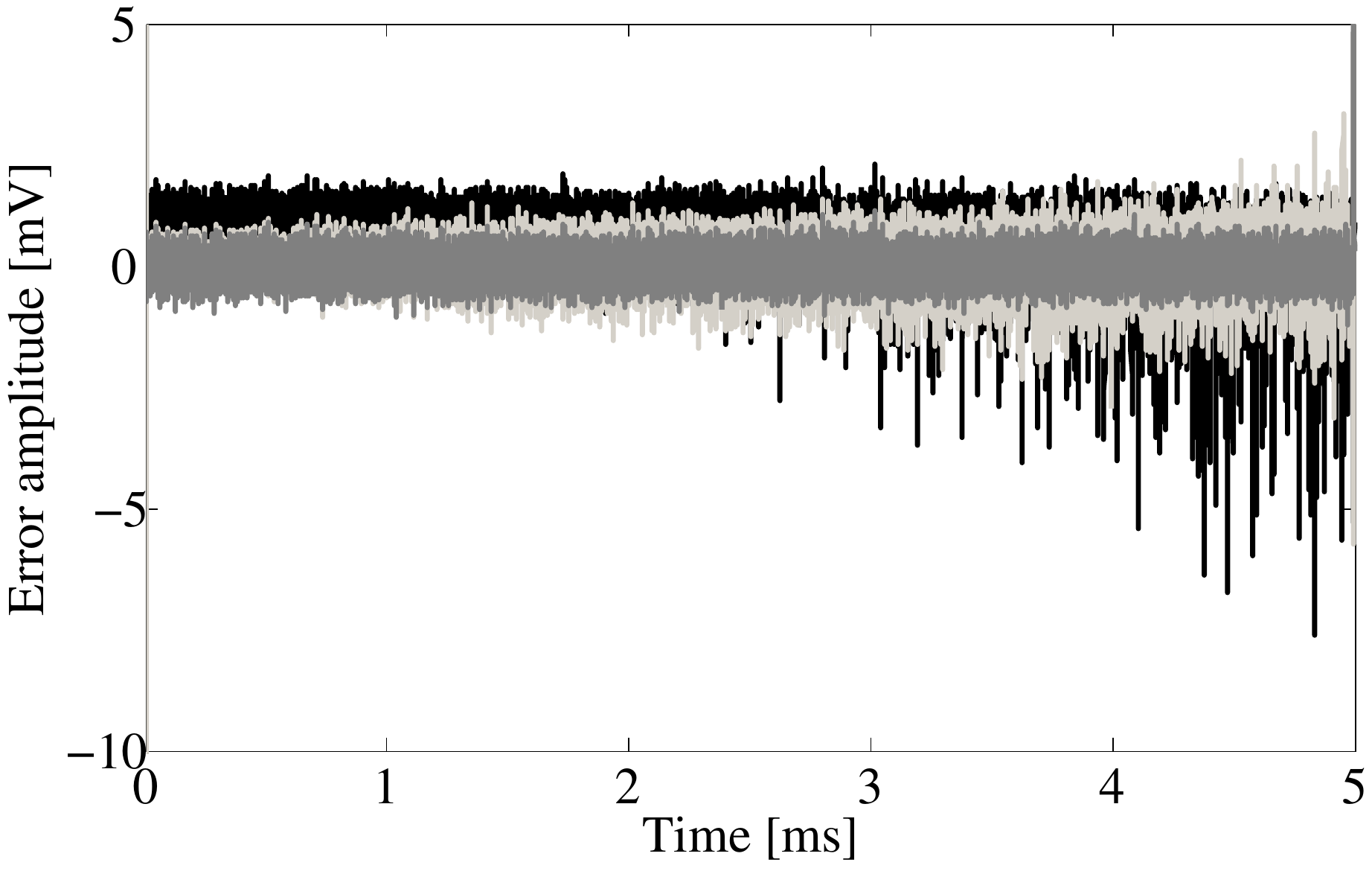}
\caption{Crystal detector example. Error (in mV) on the validation set for the BLA (black), the initialized nonlinear model (light gray) and the fitted nonlinear model after optimization (dark gray).}
\label{figerrcrystal}
\end{figure}

Fig.~\ref{figerrcrystal} shows the error obtained on the validation set by the different models. The optimized NLSS model can capture well the nonlinear behavior of the device, as indicated by the fact that the large spikes that characterize the linear model error are not present anymore in the nonlinear model error.

The results obtained by applying the proposed initialization scheme to the identification of the crystal detector can be considered very good. If compared with the error of the BLA, the initialized NLSS model gives a RMSE value on the validation data which is almost 50$\%$ lower. Moreover, starting from this initial parameter estimate, the performance of the nonlinear model is further improved in the optimization step, and the final nonlinear model error is very close to the noise level (0.23 mV). 

In the final part of this section, the obtained results are compared with the performance of other nonlinear identification methods that have been applied on the same problem.

\subsection{Comparison with other methods}

A physical description of the crystal detector is developed in \cite{Sch08}, where the internal structure is modeled as a nonlinear feedback, with a Wiener-Hammerstein branch in the feedback loop, for which the nonlinear block is a 9th order polynomial. This model (characterized by 14 parameters) results in a RMSE value of 0.30 mV. Therefore, the NLSS model obtained with the proposed algorithm performs better, and its complexity is only slightly higher (17 parameters).

The PNLSS approach \cite{pad10} has also been applied on the same problem, resulting in a RMSE of 0.26 mV (slightly better than the proposed algorithm), at the cost of a higher complexity (53 parameters).

Finally, a different method based on the identification of a nonlinear LFR block structure gives a RMSE of 0.286 mV (with 11 parameters) \cite{vmul12}.

It can be concluded that the proposed NLSS identification algorithm yields very good results in terms of both RMSE and complexity.

\section{Conclusion}
\label{sec9}

In this work a novel initialization scheme for the identification of NLSS models has been presented. The approach was successfully applied on two measurement examples.
The proposed initialization procedure has several advantages, as (i) it is a general scheme that can be used in combination with different choices of nonlinear model structures; (ii) the separation between system dynamics and nonlinear terms makes it possible to identify them independently; (iii) many nonlinear model structures can be tested rapidly on the obtained regression problem; (iv) two different fields - system identification and nonlinear regression/statistical learning - are brought together, combining the advantages of both.

A direction for future research could be to gradually impose more structure on the obtained NLSS models, e.g. by rotation of the state, to reveal the underlying structure of the considered system. Another possibility could be to start up the nonlinear initialization with a larger number of basis functions, and then select a smaller set during optimization, for instance by adding a regularization term in the cost function.

%
\section*{Acknowledgment}
This work is sponsored in part by the Fund for Scientific Research (FWO-Vlaanderen), by the Flemish Government (Methusalem),  by the Belgian Government through the Inter university Poles of Attraction (IAP VII) Program, and the ERC Advanced Grant SNLSID.
J. Suykens acknowledges support from KU Leuven, the Flemish government, FWO, the Belgian federal science policy office and the European Research Council (CoE EF/05/006, GOA MANET, IUAP DYSCO, FWO G.0377.12, ERC AdG A-DATADRIVE-B).

\bibliographystyle{IEEEtran}
\bibliography{IEEEabrv,bibtex}

\end{document}